\documentclass[sigconf,authorversion,nonacm]{acmart}

\AtBeginDocument{%
  \providecommand\BibTeX{{%
    \normalfont B\kern-0.5em{\scshape i\kern-0.25em b}\kern-0.8em\TeX}}}

\usepackage{DejaVuSans}
\usepackage{multirow}
\usepackage{subcaption}
\usepackage{arabtex}

\begin{document}

\title{SEMOUR:  A \textit{S}cripted \textit{Emo}tional Speech Repository for \textit{Ur}du }

\author{Nimra Zaheer}
\affiliation{%
 \institution{Information Technology University}
 \streetaddress{Arfa Software Technology Tower, 346-B, Ferozepur Road}
 \city{Lahore}
 \country{Pakistan}}
\email{phdcs17001@itu.edu.pk}

\author{Obaid Ullah Ahmad}
\affiliation{%
 \institution{Information Technology University}
 \streetaddress{Arfa Software Technology Tower, 346-B, Ferozepur Road}
 \city{Lahore}
 \country{Pakistan}}
\email{obaidullah.ahmad@itu.edu.pk}

\author{Ammar Ahmed}
\affiliation{%
 \institution{Information Technology University}
 \streetaddress{Arfa Software Technology Tower, 346-B, Ferozepur Road}
 \city{Lahore}
 \country{Pakistan}}
\email{ammar.ahmed@itu.edu.pk}

\author{Muhammad Shehryar Khan}
\affiliation{%
 \institution{Information Technology University}
 \streetaddress{Arfa Software Technology Tower, 346-B, Ferozepur Road}
 \city{Lahore}
 \country{Pakistan}}
\email{shehryar.khan@itu.edu.pk}

\author{Mudassir Shabbir}
\affiliation{%
 \institution{Information Technology University}
 \streetaddress{Arfa Software Technology Tower, 346-B, Ferozepur Road}
 \city{Lahore}
 \country{Pakistan}}
\email{mudassir.shabbir@itu.edu.pk }

\renewcommand{\shortauthors}{Zaheer et al.}


\begin{abstract}

  Designing reliable Speech Emotion Recognition systems is a complex task that inevitably requires sufficient data for training purposes. Such extensive datasets are currently available in only a few languages, including English, German, and Italian. In this paper, we present SEMOUR, the first scripted database of emotion-tagged speech in the Urdu language, to design an Urdu Speech Recognition System. Our gender-balanced dataset contains $15,040$ unique instances recorded by eight professional actors eliciting a syntactically complex script. The dataset is phonetically balanced, and reliably exhibits a varied set of emotions as marked by the high agreement scores among human raters in experiments. We also provide various baseline speech emotion prediction scores on the database, which could be used for various applications like personalized robot assistants, diagnosis of psychological disorders, and getting feedback from a low-tech-enabled population, etc. On a random test sample, our model correctly predicts an emotion with a state-of-the-art $92\%$ accuracy.

\end{abstract}


\begin{teaserfigure}
\includegraphics[width=\textwidth]{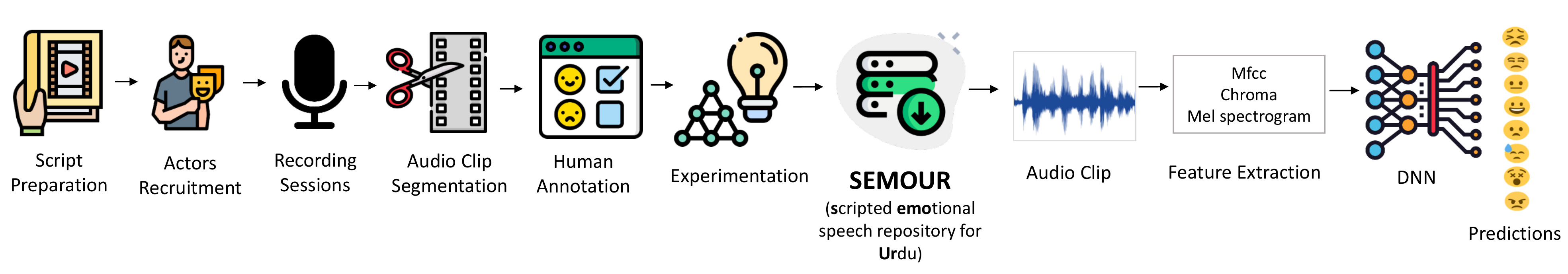}
  \caption{This figure elaborates steps involved in the design of SEMOUR: \textit{s}cripted \textit{emo}tional speech repository for \textit{Ur}du along with the architecture for our proposed machine learning network for Speech Emotion Recognition.}
  \label{main_diagram}
  \Description{The figure represents a step by step procedure to procure SEMOUR. The first stage is script preparation followed by recruitment of actors. The next step is to get the script elicited in diverse emotions by actors. Further, recorded clips are pruned according the script instance sequence, correction pronunciation and elicitation. The next step is to tag these instances by human annotators and perform experimentation. Lastly, machine learning techniques are applied on the prepared dataset.}
\end{teaserfigure}

\maketitle

\section{\label{section_intro} Introduction}

Sound waves are an information-rich medium containing multitudes of properties like pitch (relative positions of frequencies on a scale), texture (how different elements are combined), loudness (intensity of sound pressure in decibels), and duration (length of time a tone is sounded), etc. People instinctively use this information to perceive underlying emotions, e.g., sadness, fear, happiness, aggressiveness, appreciation, sarcasm, etc., even when the words carried by the sound may not be associated with these emotions. Due to the availability of vast amounts of data, and the processing power, it has become possible to detect and measure the intensity of these and other emotions in a sound bite. Speech emotion recognition systems can be used in health sciences, for example in aiding diagnosis of psychological disorders and speech-based assistance in communities and homes and elder care~\cite{weather, application_elderly,application_smarthome}. Similarly, in developing countries where it is difficult to get feedback from the low-tech-enabled population using internet-based forms, it is possible to get a general overview of their opinion by getting feedback through speech recording and recognizing the emotions in this feedback \cite{application_deep2}. Another potential use of such systems is in measuring any discrimination (based on gender, religion, or race) in public conversations \cite{abuse}.

A prerequisite for building such systems is the availability of large well-annotated datasets of sounds emotions.  Such extensive datasets are currently available in only a few languages, including English, German, and  Italian.
Urdu is the $11^{th}$ most widely spoken language in the world, with $171$ million total speakers \cite{eberhard2020ethnologue}. Moreover, it becomes the third most widely spoken language when grouped with its close variant, Hindi. Many of the applications mentioned above are quite relevant for the native speakers of the Urdu language in South Asia. Thus, a lot of people will benefit from a speech emotion recognition system for the Urdu language. In this paper, we address this research problem and present the first comprehensive emotions dataset for spoken Urdu. We make sure that our dataset approximates the common tongue in terms of the distribution of phonemes so that models trained on our dataset will be easily generalizable. Further, we collect a large dataset of more than $15,000$ utterances so that state-of-the-art data-hungry machine learning tools can be applied without causing overfitting. The utterances are recorded in a sound-proof radio studio by professional actors. Thus, the audio files that we share with the research community are of premium quality. We have also developed a basic machine learning model for speech emotion recognition and report an excellent accuracy of emotion prediction Figure~\ref{main_diagram}. In summary, the paper has the following contributions:

\begin{itemize}
  \item We study the speech emotion recognition in Urdu language and build the first comprehensive dataset that contains more than $15,000$ high-quality sound instances tagged with eight different emotions.
  \item We report the results of human accuracy of detecting emotion in this dataset, along with other statistics that were collected during an experiment of  $16$ human subjects annotating about $5,000$ utterances with an emotion.
  \item We train a basic machine learning model to recognize emotion in spoken Urdu language. We report an excellent cross-validation accuracy of $92\%$ which compares favorably with the state-of-the-art.
\end{itemize}
In the following section, we provide a brief literature overview of existing acoustic emotional repositories for different languages and Speech Emotion Recognition (SER) systems. In Section $3$ , we provide details about the design and {recording} of our dataset. The results of the human annotation of SEMOUR are discussed in Section $4$. In Section $5$, we provide details of a machine learning framework to predict emotions using this dataset for training followed by a detailed discussion in Section $6$. Finally, we conclude our work in Section $7$. 

\section{Related Work}\label{Related_work}

Prominent examples of speech databases like Emo-DB~\cite{emodb}, IEMOCAP~\cite{iemocap}, are well studied for speech emotion recognition problem. The English language encompasses multiple datasets with each one having its unique properties. IEMOCAP~\cite{iemocap} is the benchmark for speech emotion recognition systems, it consists of emotional dialogues between two actors. The dyadic nature of the database allows systems to learn in a way that it performs better when used in real life. IEMOCAP contains $12$ hours of audio-video recordings performed by $10$ actors and it has $10039$ utterances rated by three annotators. The database consists of $5$ emotions. Another well-studied database is Emo-DB~\cite{emodb} which is in the German language and it contains seven emotions. Each emotion contains almost the same number of instances making it class-balanced which means it might be easier for learning algorithms to learn the emotions distribution. There are ten sentences in seven different emotions uttered by ten different speakers with a total of $535$ instances. Emovo is an emotional corpus in the Italian language containing seven emotions performed by six actors~\cite{emovo}. The content of the corpus is semantically constant to allow the tone of the delivery to play a greater role in predicting the emotion of an instance. The same methodology is also used in~\cite{vesus} while designing a database for the English language. Other dataset for English language include MSP-IPROV \cite{mspimprov}, RAVDESS \cite{ravsdess}, SAVEE \cite{savee} and VESUS \cite{vesus}. Speech corpus, EMO-DB \cite{emodb}, VAM \cite{vam}, FAU Aibo \cite{faubaibo} for German language and EMOVO \cite{emovo} for Italian language are also available. Furthermore, CASIA \cite{casia}, CASS \cite{cass} and CHEAVD \cite{cheavd} exists for Mandarin, Keio ESD \cite{keioesd} for Japanese, RECOLA \cite{recola} for French, TURES \cite{tures} and BAUM-1 \cite{baum1} for Turkish, REGIM-TES \cite{regimtes} for Arabic and SheMo \cite{shemo} for Persian language. There are a lot of other databases in various languages along with their uses whose details can be found in~\cite{swain2018databases}.
\paragraph{Speech Emotion Recognition (SER) systems}
Speech emotion recognition systems that are built around deep neural networks are dominant in literature. They are known to produce better results than classical machine learning algorithms~\cite{fayek2017evaluating}. They are also flexible to various input features that can be extracted from speech corpus. Neural network architectures for speech emotions are built around three-block structures. First is the convolutional neural network (CNN) block to extract local features, then it is followed by a recurrent neural network (RNN) to capture context from all local features or to give attention~\cite{bahdanau2014neural} to only specific features. In the end, the weighted average is taken of the output using a fully connected neural network layer~\cite{chen20183, zhao2019speech}. It has been shown that the CNN module combined with long short-term memory (LSTM) neural networks works better than standalone CNN or LSTM based models for SER tasks in cross-corpus setting~\cite{parry2019analysis}. Other deep learning-based models like zero-shot learning, which learns using only a few labels~\cite{xu2019autonomous} and Generative Adversarial
Networks (GANs) to generate synthetic samples for robust learning have also been studied~\cite{chatziagapi2019data}.

Various features are used for solving speech emotion recognition problem. They vary from using Mel-frequency Cepstral Coefficients (MFCC)~\cite{sahidullah2012design} to using Mel-frequency spectrogram~\cite{stevens1937scale}. Other most commonly used ones are ComParE~\cite{schuller2016interspeech} and extended Geneva Minimalistic Acoustic Parameter Set (eGeMAPS)~\cite{eyben2015geneva}. 
To improve the accuracy of SER systems, additional data has been incorporated with the speech corpus. Movie script database has been used to generate a personalized profile for each speaker while classifying emotion for individual speakers ~\cite{li2019attentive}. Fusion techniques, that fuse words, and speech features to identify emotion have been studied in~\cite{sebastian2019fusion}. 

\paragraph{Resources for the Urdu Language}

The only previously available emotionally rich dataset for URDU language~\cite{urdu} consists of spontaneous speech extracted from talk shows. It contains only $400$ audio clips with 4 emotion classes and its instances lack proper sentence-level segmentation. It includes instances that have no vocals. These instances are validated by only $2$ annotators each, and no further details of validation are provided. The size of the dataset and quality of vocals in audio clips is not suited for the learning algorithm to generalize various emotions over a complex vocabulary.

 Apart from the above mentioned emotional dataset for the Urdu language, there are several large corpora of spoken Urdu developed in previous works with some applications in Speech automation. Unfortunately, none of these datasets are labeled for emotions. Also, the instances in these datasets lack the emotional aspect of the speech and monotonically fall within the natural class of emotion. Two of the most famous datasets for spoken Urdu are designed for models that would benefit from neutral or emotion-free speech \cite{raza2009design,ali2012medium}. The applications of such datasets include text-to-speech systems and automatic speech recognition systems to detect and recognize speaker attributes, for example, gender, age, and dialect. They are not suitable for a generalized speech emotion recognition system.

\section{SEMOUR Script Design}

 Dataset construction for speech instances is a tiresome endeavor with a lot of inherent challenges. In this section, we provide script design details and the obstacles faced during the process.

 Urdu is an Indo-Aryan language with $171$ million speakers all over the world according to the Ethnologue $23^{rd}$  edition \cite{eberhard2020ethnologue}. It is the national language of Pakistan and an official language in more than five states of India. The language is known for its exceptionally rich inventory of $67$ phonemes as compared to 36 phonemes of English and 35 phonemes of Chinese \cite{phonemelist}.
  It follows a Perso-Arabic written style named \textit{Nastaliq} \cite{urdufreqwords}. There are four dialects namely  Urdu, Dakhini, Hyderabadi Urdu and Rekhta \cite{dialects}. There exist six diverse types of accents namely Urdu, Punjabi, Pashto, Saraiki, Balochi, and Sindhi for Urdu dialect \cite{accents}. A language spoken by such a large community poses diverse challenges while procuring a speech dataset. Diversity in dialect is one of the major concerns and selecting speakers to cover such diversity is a laborious task. Moreover, coverage of all phonemes plays a key role in designing a rich acoustic dataset.

 As mentioned above, our focus while building this dataset is to mimic routine conversations among native speakers. Therefore, the goal of a phonetically balanced repository is to ensure that frequencies of all instances of sounds closely approximate their densities in the set of spoken words. To achieve this goal, we use two sources. Firstly, we considered the set of top $5000$ most frequently used Urdu words that appear in the list collected here \cite{urdufreqwords}.

 Secondly, we use a complete Urdu lexicon of about $46,000$ words that were collected in \cite{zia2018}.
 A uniformly random sample of words from any one of these sources will have, in expectation, the property of being phonetically balanced. However, we also wanted our dataset to have phrases and sentences with sufficient lexical complexity.
 Our script consists of $235$ instances composed of $43$ common words , $66$ two-word phrases, and $126$ simple-structured Urdu sentences as shown in Table \ref{stats}.
 Frequently used vocabulary, and a small subset of emotionally enabled sentences against each emotion has also been incorporated in the preparation of the script.

\begin{figure*}[t]
  \centering
  \includegraphics[scale=0.45]{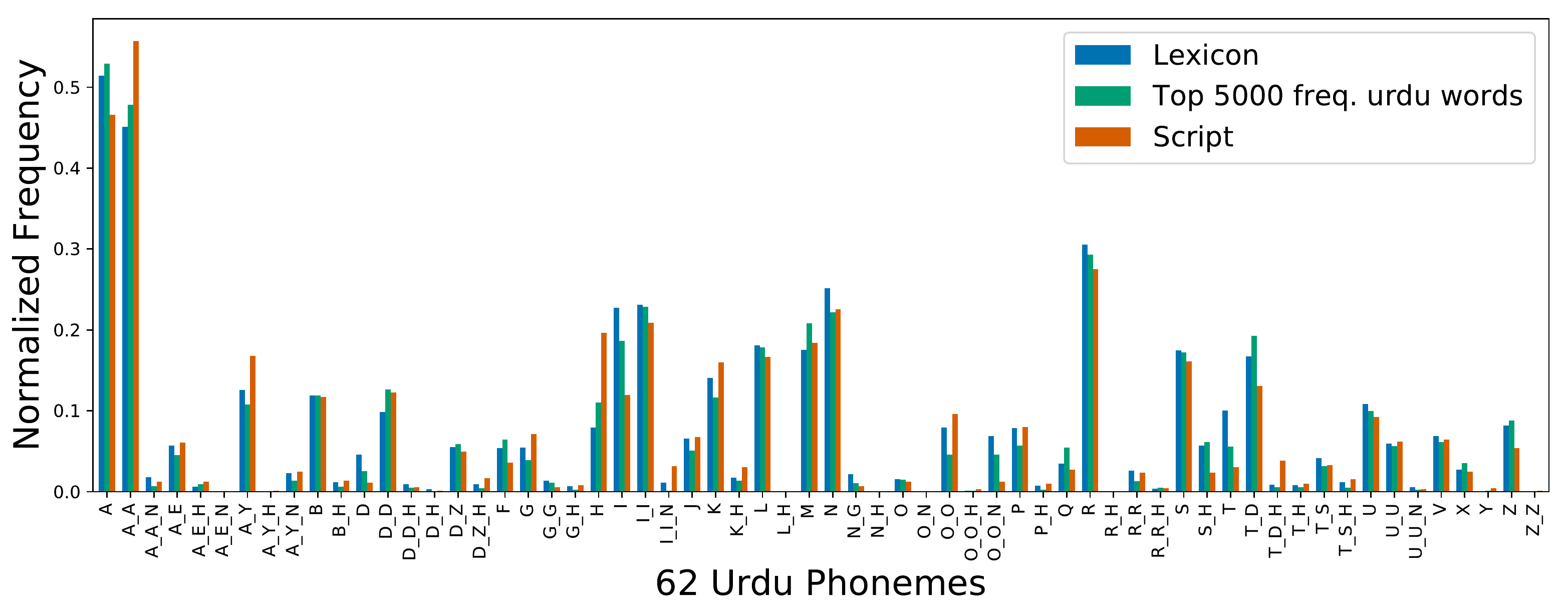}
  \caption{Phoneme comparison of our designed script with Urdu lexicon of $46,000$ words, and $5000$ most frequently used words \cite{zia2018}. On the $x$-axis, we have a list of $62$ phonemes, and on $y$-axis we have normalized phoneme occurrence frequency. The histogram shows that SEMOUR is phonetically balanced when compared to the other two standard datasets. Normalized frequency scale ranges from $0$ to $1$.}
  \label{phoneme_analysis}
  \Description{This figure elaborates comparison between Urdu language phonemes using bars of normalized frequency of words of our design script, rich Urdu word Lexicon and 5000 frequently used Urdu words respectively. A certain balance is  observed among various phonemes indicating that our script is phonetically balanced.}
  \end{figure*}

 We used methods in \cite{zia2018}, to compute phoneme frequency for our script and compared it against the two sources as shown in Figure \ref{phoneme_analysis}. It can be seen that our script covers all phonemes as frequently as they appear in the two standard datasets almost always. Also note that model \cite{zia2018} is trained for $62$ phonemes, hence Figure \ref{phoneme_analysis} shows plot for phoneme normalized frequencies against $62$ phonemes.

\begin{table*}[t]
\caption{Utterance and script level statistics for SEMOUR.}

\label{stats}
\begin{tabular}{c  c  @{\hskip 0.7in} c  c}
  \toprule
\textbf{Category} &\textbf{Instances}&  \textbf{Category}        & \textbf{Statistics}           \\ 
 \midrule
Words             & 43     & Avg. time per instance   & 1.657 sec                     \\ 
Phrases           & 66 & Avg. time per Actor      & 51 min, 56 sec                \\ 
Sentences         & 126   & Avg. time per Emotion    & 6 min, 29 sec                 \\ 
\textbf{Total instances}   & \textbf{235} &\textbf{Total instances} & \textbf{15,040}               \\ 
\textbf{Unique words}      & \textbf{559}      &\textbf{Total duration}  & \textbf{6 hr, 55 min, 29 sec} \\ 
  \bottomrule
  \Description{This table shows number of words, phrases and sentences in our script followed by average time taken per instance, actor and emotions.}
\end{tabular}
\end{table*}

We used the services of a local radio studio to conduct soundproof recording sessions. 
After the recording, basic noise reduction, voice normalization, and amplification techniques are applied to the audio clips. Each recording is manually segmented based on script sequence, and emotion instance. The final dataset consists of more than $15,000$ utterances with a cumulative duration of \textbf{6 hr, 55 min, 29 sec}. We have uploaded high and low definition versions of our dataset\footnote{\url{http://acoustics-lab.itu.edu.pk/semour/}} and are making it publicly available for the research community. Each audio clip in the low definition version of the dataset has $2$ channels (stereo) with sample rate, sample size, and bit rate of $44.100$ kHz, $16$ bit, and $1411$ kbps, respectively. Each audio clip in high definition has $2$ channels (stereo) with sample rate, sample size, and bit rate of $48.000$ kHz, $24$ bit, and $2304$ kbps, respectively. Table \ref{stats} shows the details of utterance and script level statistics.

\section{Human Evaluation of SEMOUR}
\label{section_human_evaluation}
The first experiment that we performed on SEMOUR, was the human evaluation of the dataset. We selected one-third of the audio clips uniformly at random from our dataset and presented them to the evaluators one at a time with a list of questions. We used sixteen evaluators to annotate about $5,000$ clips with each clip receiving at least two evaluations. 
All of the annotators have proficient Urdu speaking and Listening skills. All but two have Urdu as their first language; the other two have Urdu as a second language. All the evaluators have a minimum of secondary school education aging from $17$ to $32$ years old.

After listening to each clip, evaluators were asked to input two evaluations, namely \textit{discrete categorical} (a choice among neutral, happiness, surprise, sadness, boredom, fearful, anger, and disgust) and \textit{continuous attribute} (a numeric value for each of valence, dominance, and activation). Valence encapsulates degrees of pleasantness elicited in an audio clip varying from negative to positive, activation depicts the levels of intensity such as calmness or excitement and dominance portrays the speaker’s control over the utterance i.e. weak or strong \cite{RUSSELL1977273}. Furthermore, an actor's performance was also rated in terms of \textit{genuineness}, i.e., fake or natural. The evaluators were given a short training on the use of the survey application and the meaning of each term used in the questions.

We performed extensive analyses on the feedback received from evaluators. As shown in Table \ref{all_votes}, we see an average accuracy of $78\%$, i.e., on average, an evaluator correctly identified the emotion in a clip with $78\%$ accuracy. With majority voting, we observe an average accuracy of $79\%$ as shown in Table~\ref{maj_votes}.  This is a very high accuracy compared to a random classification which would result in an accuracy of $12.5\%$. One can conclude, that most of the audio files are labeled with correct emotion. Secondly, this also shows that humans can correctly perceive labeled emotions in most of these clips. However, we observe some variance in the accurate recognition of different emotions. For example, two emotions, Disgust and Fearful, were identified with the lowest accuracy as compared to other emotions. It turns out that some instances of Disgust were confused with Anger and Neutral emotions. We suspect that this emotion is very hard to differentiate from other emotions in general, and probably, not as commonly used as other emotions. Similarly, Fear was incorrectly classified as Sadness in some instances because shivers while crying were perceived as hiccups.

\begin{figure*}
\centering
\subfloat[]{\includegraphics[width=.4\textwidth]{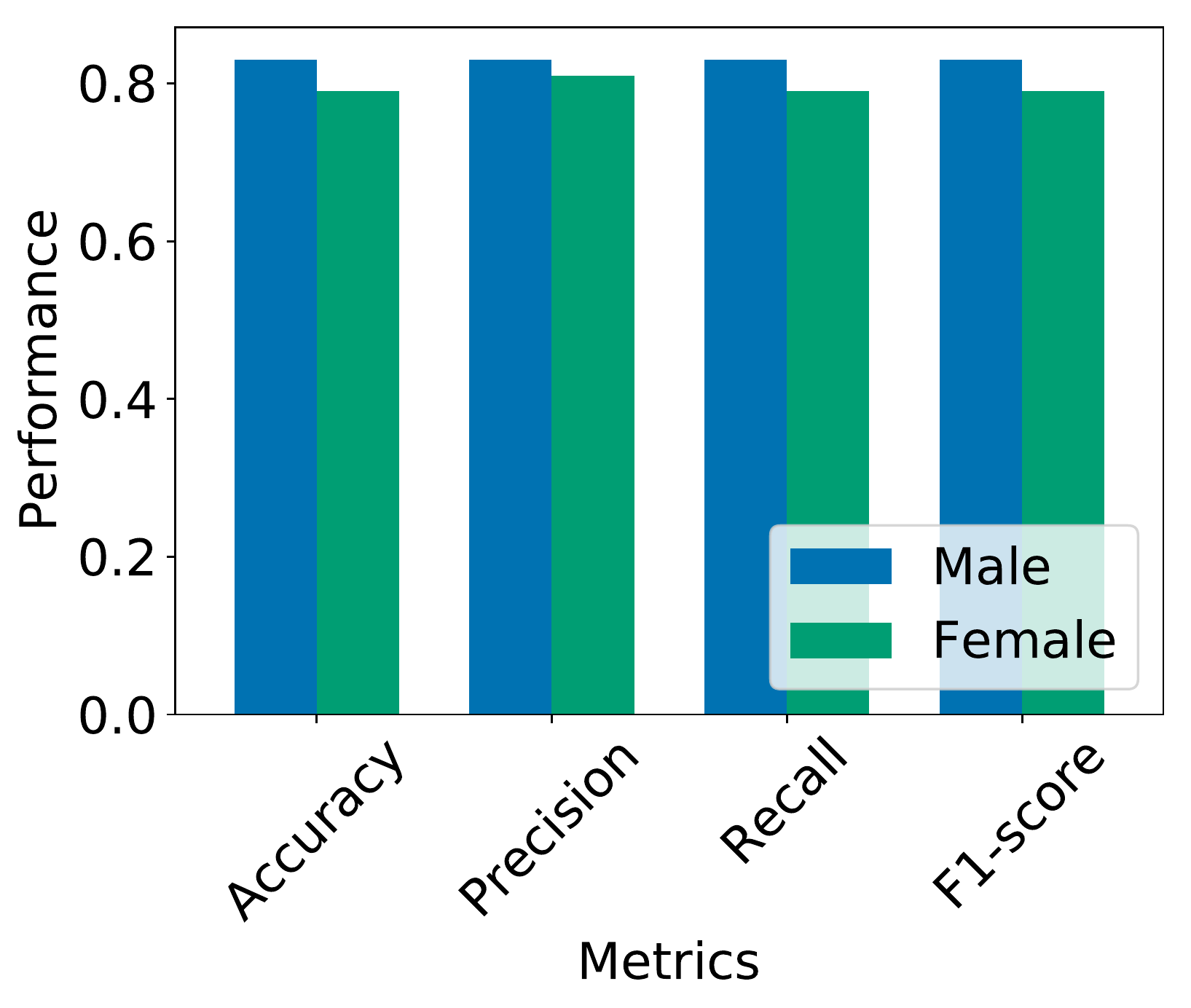}%
\label{gen_analysis}
}
\hfil
\subfloat[]{\includegraphics[width=.4\textwidth]{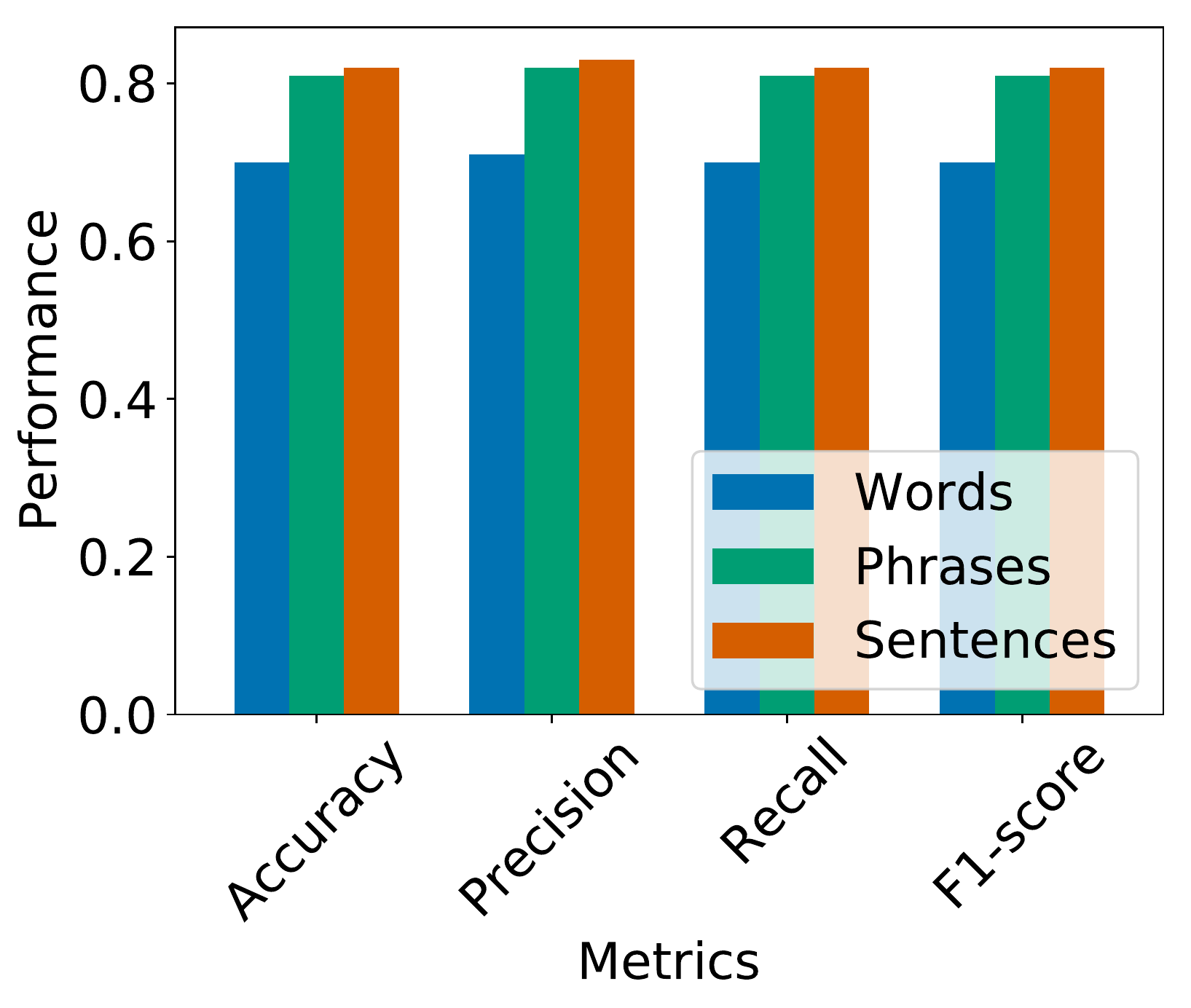}%
	 \label{syn_analysis}
}
\hfil
\subfloat[]{\includegraphics[width=.5\textwidth]{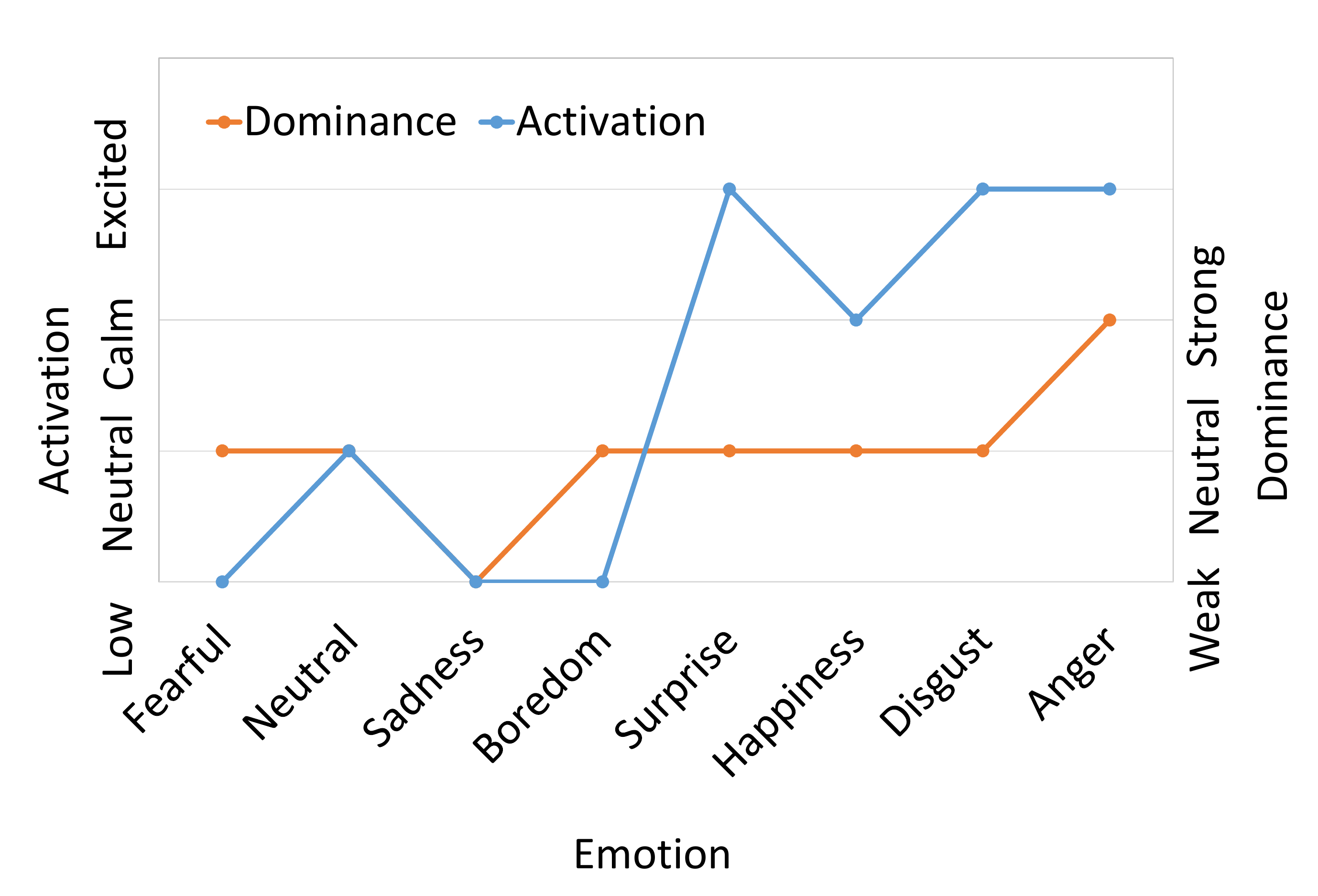}%
 \label{cont_analysis}
}
\caption{Results for experimentation on human annotation : (a) performance analysis against evaluation metrics for male and female actors. Male actors perform better than female actors when their clips were annotated. The values against males for accuracy, recall, precision, F1 score are $0.83, 0.83, 0.83$ and $0.83$ respectively. The values against females for accuracy, recall, precision, F1 score are $0.79, 0.81, 0.79$ and $0.79$ respectively. (b) performance analysis against evaluation metrics for the script's lexical complexity. Sentences cover rich emotion than phrases and so on. The values against words for accuracy, recall, precision, F1 score are $0.70, 0.71, 0.70$ and $0.70$ respectively. The values against phrases for accuracy, recall, precision, F1 score are $0.81, 0.82, 0.81$ and $0.81$ respectively.The values against sentences for accuracy, recall, precision, F1 score are $0.82, 0.83, 0.82$ and $0.82$ respectively. Performance scale for (a) and (b) ranges from $0$ to $1$. (c) Continuous attributes categorization for each emotion. The left and right y-axis depict activation (low, natural, calm, and excited) and dominance (weak, neutral, and strong) respectively. Surprise has excited activation and neutral dominance.}
\label{ann_exp}
\Description{Figure (a) elaborates a bar chart indicating gender based analysis on emotion elicitation performance in terms of accuracy, precision, recall and F1-score. Figure (b) elaborates a bar chart indicating syntactic complexity analysis on emotion elicitation performance in terms of accuracy, precision, recall and F1-score. Figure (c) is a line chart for continuous attributes against emotion having activation and dominance on left and right x-axis respectively. }
\end{figure*}

Furthermore, these ratings also give us a yardstick to measure the performance of individual actors. 
For each actor, accuracy, precision, recall, f1-score along with Cohen's kappa score for measuring rater's agreement are presented in Table \ref{per_actors}. It can be observed that actor number $8$ performed well and actor number $3$ has the highest rater's agreement. 

Moreover, all scores are greater than 0.4, (i.e., fair agreement), we conclude that our actors performed well and the general audience was able to distinguish among the emotions uttered. Additionally, while measuring genuineness of acting, there were 115 and 267 votes for \textit{fake} category for males and females, respectively, implying that male actors may have performed relatively better as shown in Figure \ref{gen_analysis}. 
\begin{table*}
\small
\caption{Individual performance of actors with majority voting}
\label{per_actors}

\begin{tabular}{c @{\hskip 0.5in} c c c c c}
\toprule
\textbf{Actors} & \textbf{Accuracy} & \textbf{Precision} & \textbf{Recall} & \textbf{F1-Measure} & \textbf{Inter-evaluator score} \\ \midrule
1               & 0.66              & 0.68               & 0.66            & 0.65                & 0.54                           \\ 
2               & 0.87              & 0.88               & 0.87            & 0.87                & 0.71                           \\
3               & 0.79              & 0.8                & 0.79            & 0.79                & \textbf{0.89}                  \\ 
4               & 0.86              & 0.86               & 0.86            & 0.86                & 0.67                           \\ 
5               & 0.82              & 0.83               & 0.82            & 0.81                & 0.73                           \\ 
6               & 0.8               & 0.81               & 0.8             & 0.79                & 0.71                           \\
7               & 0.66              & 0.72               & 0.66            & 0.64                & 0.42                           \\ 
8               & \textbf{0.91}     & \textbf{0.92}      & \textbf{0.91}   & \textbf{0.91}       & 0.88                           \\ \bottomrule
\Description{This table elaborates actor wise performance metrics followed by inter-evaluator score highlighting individual performance of actors. }
\end{tabular}
\end{table*}

We have also performed experiments based on the syntactical complexity of our script, as naturally perceived single words and phrases encapsulate fewer emotions than sentences. Intuitively, it should be harder to utter emotion in a word or phrase. Our results confirm this intuition in terms of accuracy, precision, recall, and f1-measure as shown in Figure \ref{syn_analysis}. Along with discrete labels, raters were asked to rank activation and dominance for each utterance. With majority voting, our repository conforms with generally perceived notions of these values in the discrete emotions, e.g., sadness has \textit{low} activation and \textit{weak} dominance as compared to Anger. The results are as shown in Figure \ref{cont_analysis}. 

\begin{figure*}
\centering
\subfloat[]{\includegraphics[width=.48\textwidth]{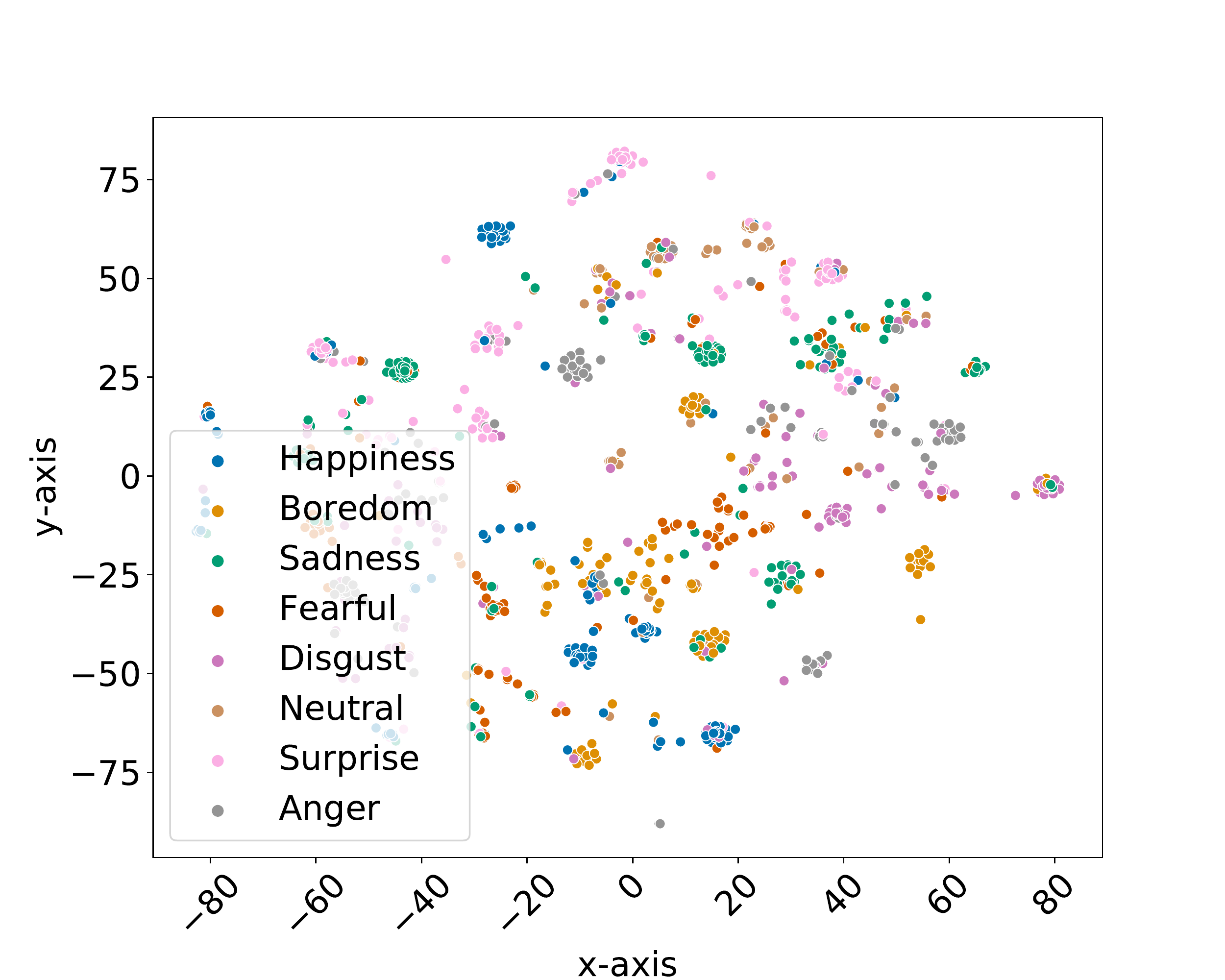}%
\label{tnse_all}
}
\hfil
\subfloat[]{\includegraphics[width=.48\textwidth]{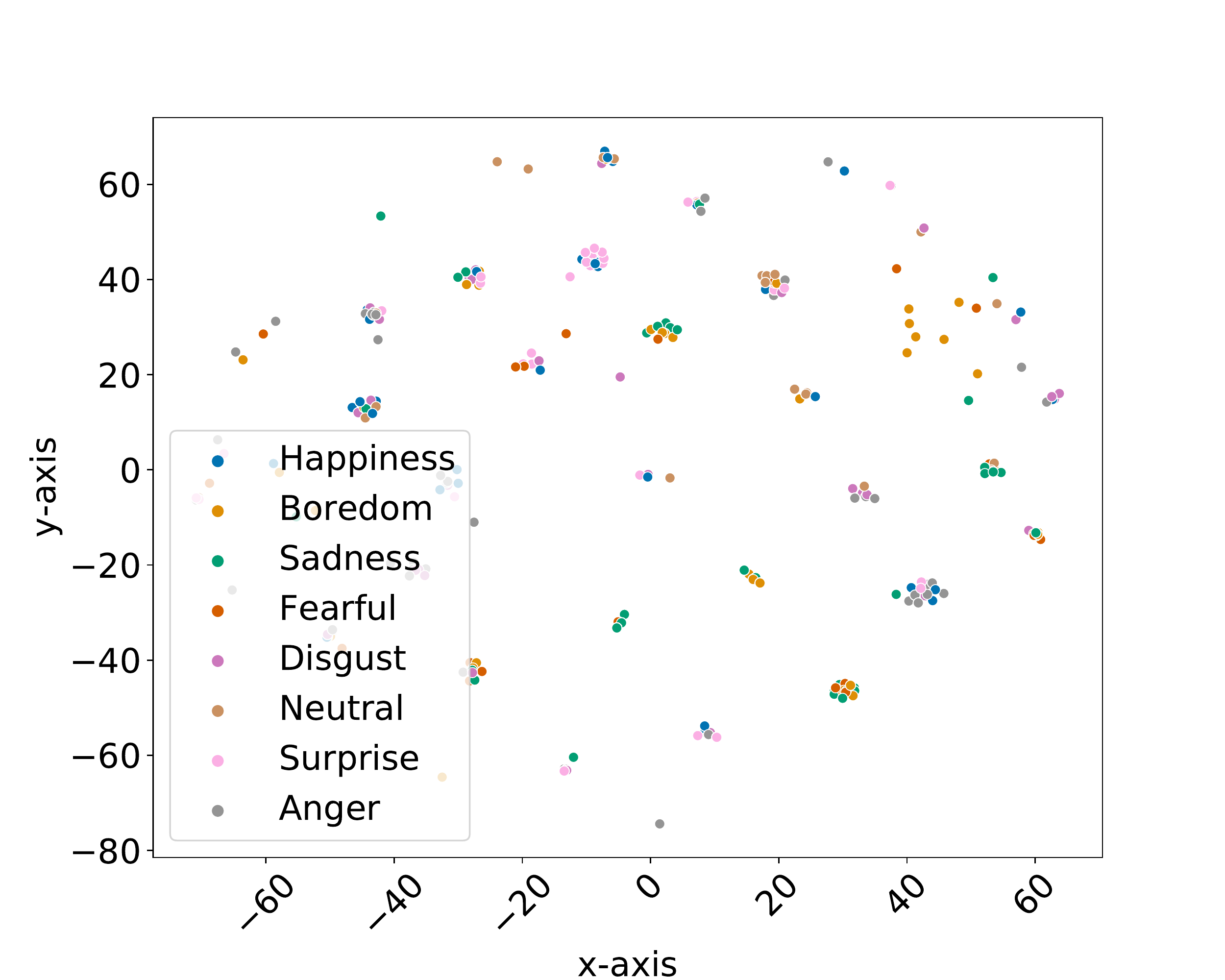}%
	 \label{tsne_rater}
}

\caption{t-SNE plots against human annotation. (a) Ground truth plotted against Tag, Valence, Activation, Dominance, Genuineness (TVADG). (b) Ground Truth plotted against Valence, Activation, Dominance, Genuineness (VADG). }
\label{tnse_ann_Exp}
\Description{The figure is a t-SNE visualization using a scatter plots where points in graph belongs to different emotion classes.}
\end{figure*}

Lastly, we applied \textit{t-Distributed Stochastic Neighbor Embedding (t-SNE)}, a dimensionality reduction technique to our ratings for detailed analysis of emotions. Figure \ref{tnse_all} shows the t-SNE relation between ground truth and the independent variables annotator had control on, i.e., Emotion-Tag, Valence, Activation, Dominance, and Genuineness (TVADG). The distribution for various emotions in the figure shows the relative impacts of all of the annotated properties, and an overlap between two color classes indicates a potentially incorrect annotation. For example, it seems that point clouds for  Happiness and Surprise emotions are close to each other, implying that they may have been misclassified for each other. Similarly, Sadness and Disgust classes have some overlap too. On the other hand, classes of Anger, Neutral, Boredom, and Disgust emotion seem quite distinguishable from each other. Figure \ref{tsne_rater} is the t-SNE relation between ground truth and the independent variables annotator had excluding the emotion-tag property, i.e., VADG. Since the properties have a very small distribution, the clusters of different classes are subsumed into one another. Although, boredom class remains the most distinguishable one.

\begin{table*}
\small
\caption{Confusion matrix for intended vs. perceived labels with all ratings}
\label{all_votes}
\begin{tabular}{ c c c c c c c c c c c }
\toprule
\multicolumn{2}{c}{\multirow{2}{*}{\textbf{Avg. Acc. =78\%}}} & \multicolumn{9}{c}{\textbf{Perceived Labels}}                                                                                                                          \\ 
\multicolumn{2}{c}{}                                          & \textbf{Anger} & \textbf{Boredom} & \textbf{Disgust} & \textbf{Fearful} & \textbf{Happiness} & \textbf{Neutral} & \textbf{Sadness} & \textbf{Surprise} & \textbf{Other} \\ 
\midrule
\multirow{8}{*}{\textbf{\rotatebox{90}{Ground truth}}}   & \textbf{Anger}       & \textbf{972}   & 3                & 46               & 5                & 13                 & 52               & 0                & 142               & 14             \\  
                                         & \textbf{Boredom}     & 1              & \textbf{1101}    & 21               & 5                & 2                  & 65               & 45               & 0                 & 7              \\ 
                                         & \textbf{Disgust}     & 167            & 30               & \textbf{770}     & 29               & 34                 & 127              & 15               & 55                & 25             \\ 
                                         & \textbf{Fearful}     & 1              & 22               & 26               & \textbf{781}     & 12                 & 70               & 246              & 73                & 15             \\ 
                                         & \textbf{Happiness}   & 1              & 5                & 5                & 7                & \textbf{1067}      & 64               & 8                & 78                & 13             \\ 
                                         & \textbf{Neutral}     & 10             & 84               & 30               & 5                & 7                  & \textbf{1076}    & 12               & 8                 & 16             \\  
                                         & \textbf{Sadness}     & 0              & 49               & 7                & 149              & 12                 & 25               & \textbf{995}     & 6                 & 1              \\ 
                                         & \textbf{Surprise}    & 34             & 2                & 28               & 15               & 42                 & 85               & 2                & \textbf{1021}     & 16             \\ \bottomrule
\end{tabular}
\Description{The table elaborates votes extracted from human annotations without majority voting against each emotion giving an idea about distribution of votes for each emotion tag. Each cell contains a numeric value against perceived and golden labels.   }
\end{table*}


\begin{table*}
\small
\caption{Confusion matrix for intended vs. perceived labels with majority voting}
\label{maj_votes}
\begin{tabular}{c c c c c c c c c c c}
\toprule
\multicolumn{2}{c}{\multirow{2}{*}{\textbf{Avg. Acc. =79\%}}} & \multicolumn{9}{c}{\textbf{Perceived Labels}}                                                                                                                          \\ 
\multicolumn{2}{c}{}                                          & \textbf{Anger} & \textbf{Boredom} & \textbf{Disgust} & \textbf{Fearful} & \textbf{Happiness} & \textbf{Neutral} & \textbf{Sadness} & \textbf{Surprise} & \textbf{Other} \\ \midrule
\multirow{8}{*}{\textbf{\rotatebox{90}{Ground truth}}}   & \textbf{Anger}       & \textbf{542}   & 1                & 11               & 2                & 6                  & 18               & 0                & 42                & 1              \\ 
                                         & \textbf{Boredom}     & 1              & \textbf{605}     & 2                & 2                & 0                  & 11               & 2                & 0                 & 0              \\ 
                                         & \textbf{Disgust}     & 141            & 25               & \textbf{388}     & 6                & 6                  & 45               & 1                & 9                 & 4              \\ 
                                         & \textbf{Fearful}     & 1              & 21               & 22               & \textbf{473}     & 3                  & 27               & 56               & 15                & 4              \\  
                                         & \textbf{Happiness}   & 1              & 5                & 3                & 6                & \textbf{572}       & 15               & 1                & 18                & 3              \\  
                                         & \textbf{Neutral}     & 10             & 78               & 23               & 5                & 7                  & \textbf{498}     & 2                & 0                 & 1              \\  
                                         & \textbf{Sadness}     & 0              & 45               & 7                & 120              & 11                 & 18               & \textbf{418}     & 0                 & 1              \\ 
                                         & \textbf{Surprise}    & 31             & 2                & 19               & 11               & 30                 & 63               & 1                & \textbf{451}      & 13             \\ \bottomrule
\Description{The table elaborates votes extracted from human annotations with majority voting against each emotion giving an idea about distribution of votes for each emotion tag. Each cell contains a numeric value against perceived and golden labels.   }
\end{tabular}

\end{table*}

\section{Speech Emotion Recognition using Machine Learning}

The ultimate goal of building a large balanced dataset is to be able to train machine learning models that can predict emotions in an unseen sound clip. In this section, we discuss the performance, and evaluation of the first machine learning model to use SEMOUR as a dataset.  We aim to solve a multi-label classification problem that predicts emotion based on features extracted from an acoustic clip. Formally, Let $U=\{u_1,u_2,\ldots, u_n \} \in \mathbb{R}^{n \times d}$ be a list input vector where $n$ is the number of utterances in a training dataset, and $d$ is the number of features extracted from each utterance. Let output vector be $O=\{o_1, o_2, \ldots, o_n:  1\le o_i\le m \} $ where each $o_i$ is the class label associated with the training input vector $u_i$, and $m$ is the number of emotions.  Our goal is to learn a function $\mathcal{F}:U\rightarrow O$ so that, $(1). $ function $\mathcal{F}$ correctly maps emotion label $o_i$ for a feature vector $u_i$ for maximum number of instances in $U$, (training accuracy), $(2)$ $\mathcal{F}$ outputs a correct emotion label for a feature vector $u_j\notin U$ that corresponds to an unseen audio clip, assuming unseen clip is drawn from the distribution of $U$, (test accuracy).

\begin{figure*}[t]
  \centering
  \includegraphics[scale=0.7]{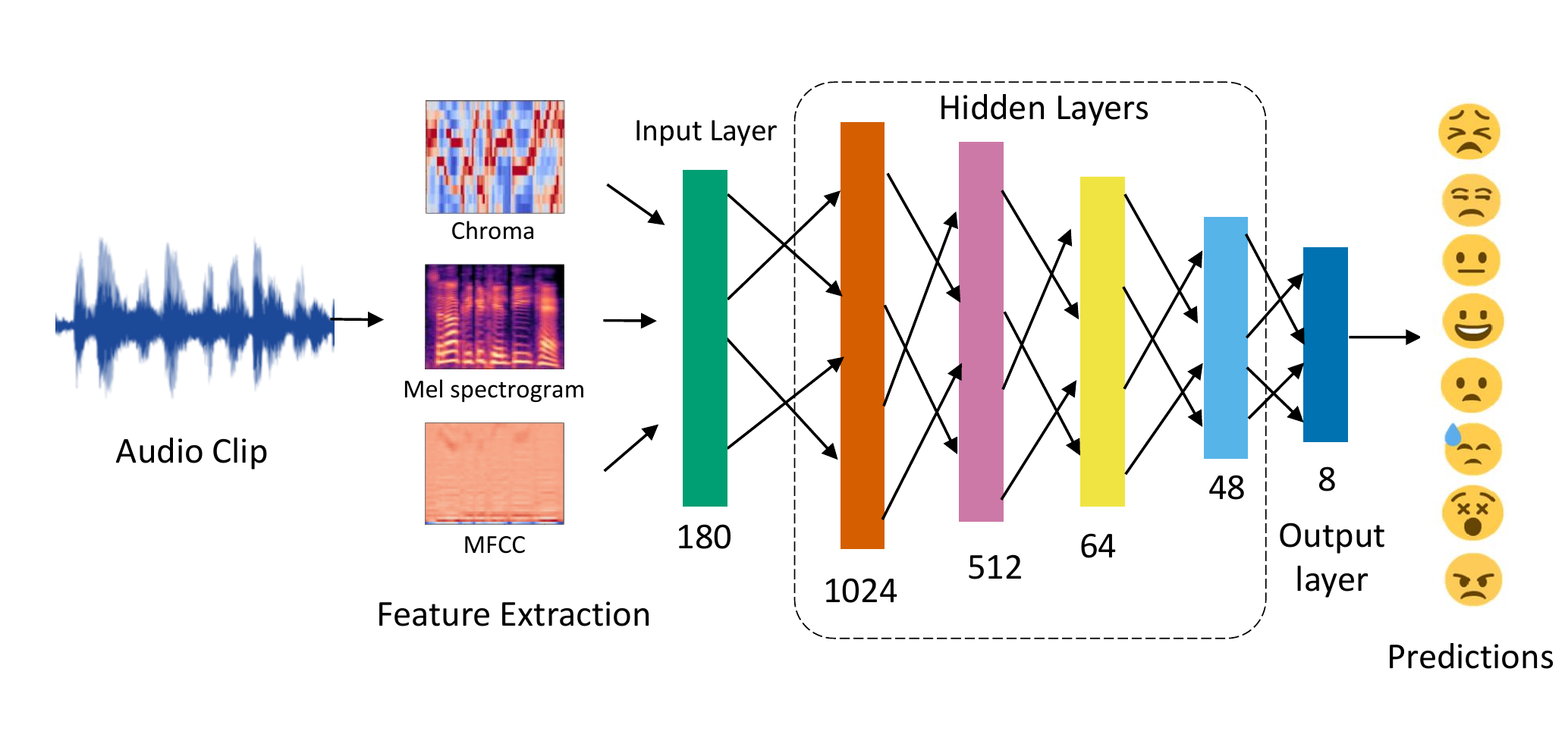}
   \caption{Proposed 5-layered neural network for Speech Emotion Recognition (SER). Three core features namely MFCCs, chromagram, and Mel-spectrogram are extracted from each audio clip and fed to a 5- dense-layered neural network to predict 8 complex emotions.}
  \label{architecture}
\Description{This figure elaborates the architecture of our proposed model where features namely chroma, mel spectrogram andd MFCC are extracted from an audio clip, which are then fed to five layered neural network to predict eight emotions. }
\end{figure*}

The features used for training were Mel-frequency Cepstral Coefficients (MFCCs), chromagram, and Mel-spectrogram. Collectively, for each sample, an array of $40$ coefficients of MFCC, $12$ pitch classes values, and $128$ mean values of Mel-spectrogram form a feature vector of size, $d=180$. This feature vector is used for the task of classification. Visualization of features for neutral emotion audio sample are shown in Figure~\ref{architecture}.

A five-layered neural architecture was trained for classification, as shown in Figure \ref{architecture}. Four hidden layers with $1024$, $512$, $64$, and $48$ neurons were used respectively along with $8$-dimensional fully connected output layer. The parameters for epochs, L1-regularization, learning rate, batch size were set to $30$,$0.001$, $0.001$, and $50$, respectively, along with \textit{Scaled Exponential Linear Units (SELU)} and \textit{softmax} as activation functions for hidden and output layers respectively. The above mentioned three features were extracted using the Librosa speech library and were used as an input to our proposed architecture \cite{mcfee2015librosa}. 
Three different variations of the experiment namely \textit{random splits, leave one speaker out}, and \textit{gender based setting} were selected for detailed analysis. 

Our proposed architecture outperforms existing machine learning techniques as shown in Table \ref{comparison}. The comparison features of SEMOUR and the accuracy of our model are compared with the existing dataset in Table~\ref{dataset_comp}. In the following subsections, we provide comprehensive details regarding experimentation.
\begin{table*}
\caption{Result comparison with classical machine learning algorithms and our proposed deep neural network for 80\% training and 20\% testing  random split.}
\label{comparison}
\begin{tabular}{c c c c c}
\toprule
\textbf{Techniques/Evaluation Metrics} & \textbf{Accuracy} & \textbf{Precision} & \textbf{Recall} & \textbf{F1- score} \\ \midrule
Gaussian Naive Bayes                   & 0.44              & 0.45               & 0.44            & 0.41               \\ 
Logistic Regression                    & 0.64              & 0.64               & 0.64            & 0.64               \\ 
SVM                                    & 0.64              & 0.64               & 0.64            & 0.64               \\ 
Decision Tree                          & 0.69              & 0.69               & 0.69            & 0.69               \\ 
ANN                                    & 0.86              & 0.88               & 0.86            & 0.86               \\ 
Random Forest                          & 0.87              & 0.87               & 0.87            & 0.87               \\ 
Our Method                             & \textbf{0.91}     & \textbf{0.91}      & \textbf{0.91}   & \textbf{0.91}      \\ \bottomrule
\Description{This table elaborate comparison between classical machine learning algorithms with our proposed model where performance metrics are shown against each technique used.}
\end{tabular}
\end{table*}

\begin{table*}
\small
\caption{Testing our proposed architecture and dataset on \cite{urdu} by taking 80\% of SEMOUR dataset for training and 20\% for validation.}
\label{dataset_comp}
\begin{tabular}{c c c c c c c}
\toprule
\textbf{Dataset}                               & \textbf{No. of emotions} & \textbf{No. of instances} & \textbf{Accuracy} & \textbf{Precision} & \textbf{Recall} & \textbf{F1- score} \\ \midrule
Existing dataset  \cite{urdu} & 4               & 400                       & 0.21              & 0.3                & 0.21            & 0.23               \\ 
\textbf{SEMOUR}                                  & \textbf{8}     & \textbf{15,040}           & \textbf{0.91}     & \textbf{0.91}      & \textbf{0.91}   & \textbf{0.91}      \\ \bottomrule
\Description{This table elaborates SEMOUR and existing Urdu emotional repository testing results in terms of number of emotions, instances and performance metrics.}
\end{tabular}
\end{table*}

\subsection{Stochastic Division of Train and Test sets}
\begin{figure*}
    \centering
\subfloat[]{\includegraphics[width=.42\textwidth]{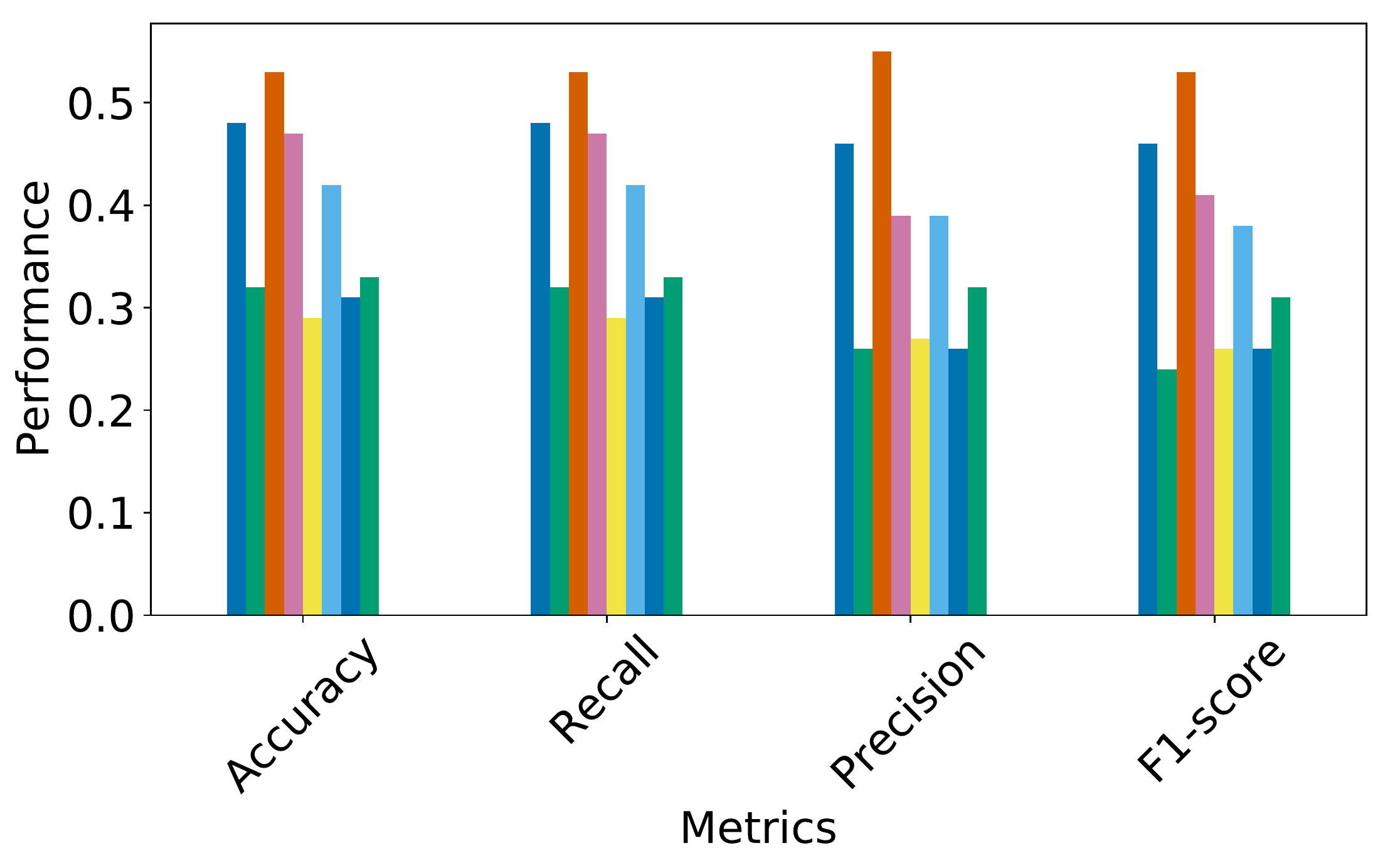}%

}
\hfil
\subfloat[]{\includegraphics[width=.5\textwidth]{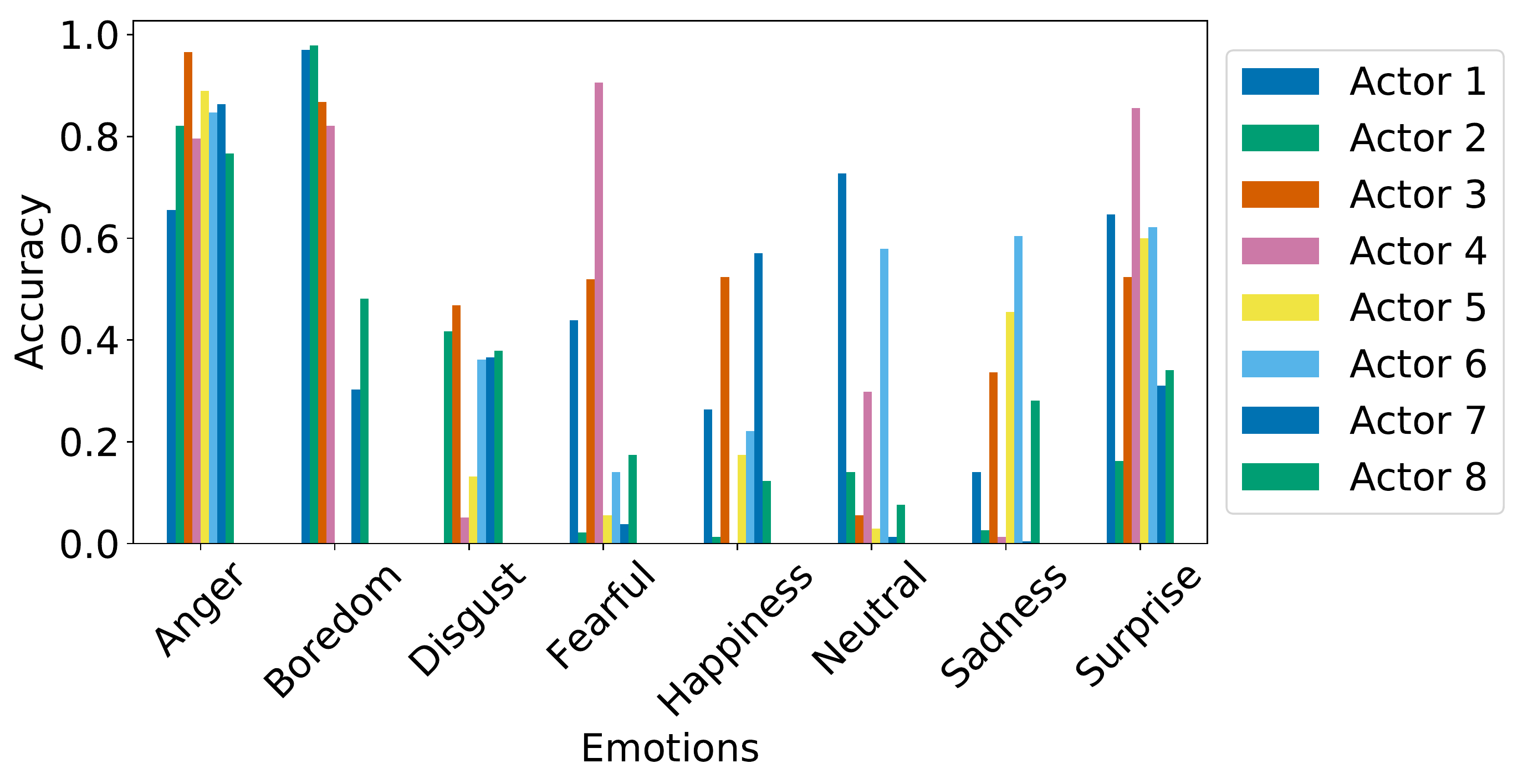}%

}
\hfil

\subfloat[]{\includegraphics[width=0.45\textwidth]{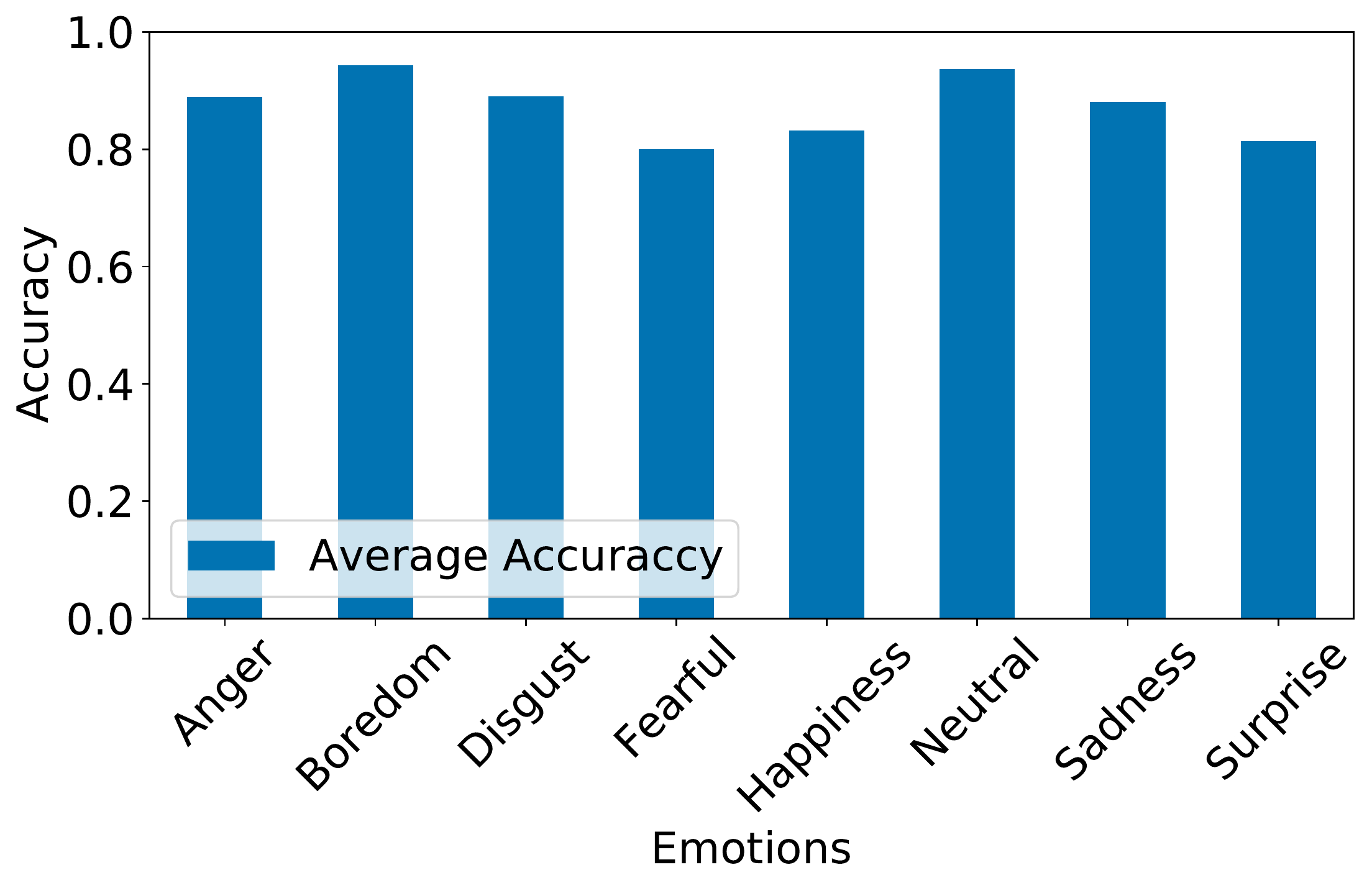}%

}
    \caption{Summarized results for SER experiments with performance ranging from $0$ to $1$. (a) Performance analysis across actors against evaluation metrics. The mean value for accuracy, recall, precision, F1 score is $0.39, 0.39, 0.36$ and $0.36$ respectively. The standard deviation for accuracy, recall, precision, F1 score is $0.09, 0.09, 0.1$ and $0.1$ respectively. (b) Performance analysis across all emotions against each actor. The mean value for anger, boredom, disgust, fearful, happiness, neutral, sadness and surprise  is $0.58, 0.52, 0.27, 0.27, 0.28, 0.24, 0.24, 0.23$ and $0.50$ respectively. The standard deviation for anger, boredom, disgust, fearful, happiness, neutral, sadness and surprise is $0.09, 0.41, 0.18, 0.31,0.21, 0.27, 0.22$ and $0.22$ respectively. The legend for (a) and (b) is shown on the right side. (c) Average accuracy against all emotions for stochastic testing. The graph shows that average accuracy is not only very high, it is also stable and consistent among all emotions. The mean value for anger, boredom, disgust, fearful, happiness, neutral, sadness and surprise  is $0.89, 0.92, 0.89, 0.80, 0.84, 0.91, 0.89, 0.84$ and $0.89$ respectively. The standard deviation for anger, boredom, disgust, fearful, happiness, neutral, sadness and surprise is $0.03, 0.04, 0.04, 0.08,0.06, 0.02, 0.06$ and $0.05$ respectively.}
    \label{all_ser}
    \Description{The figure elaborates bar charts for our experiments performed on SEMOUR. Figure (a) and (b) shows bars for each actors against performance metrics and emotions respectively. Figure (c) elaborates emotion class accuracy for stochastic testing.}
\end{figure*}

The first experiment, we designed, was based on a stochastic split experiment, where we test our accuracy on the complete dataset. For this experiment, we used the architecture as explained in Figure \ref{architecture}. We trained our model for 100 epochs on a randomly selected $90\%$ dataset ($13,536$ instances). Once trained, we test it on the remaining $1504$ instances and obtained the highest accuracy of $92\%$ and an average accuracy of $90\%$. To validate the experiment, $10$-folds cross-validation technique was used. The results are visualized in Figure~\ref{all_ser}c.
Accuracy for individual emotion class was also analyzed for the variance. We observed that our model performed exceptionally well on the Boredom and Neutral emotions with an accuracy of $99\%$, and $98\%$, respectively. The worse performing emotions were Fearful and Happiness which were identified with an accuracy of $86\%$, and $85\%$, respectively. 

\subsection{Gender-based Analysis among Actors}
The second experiment, we conducted, was a gender-based experiment where we had a binary combinatorial testing technique. We have an equal distribution of male and female actors in our dataset, therefore, in a random setting, one would expect a balancing classification accuracy. All four binary combinations were evaluated, i.e., binary choices are male and female actors, and training and testing samples.

As shown in Table~\ref{our_results} the experiments on the same gender yielded excellent results whereas the cross-gender study resulted in significantly lower accuracy. We believe this is since each speaker primarily has an independent distribution that is not concerning a specific gender, rather unique to each individual. To cement this we conducted a leave-one-out experiment on the same gender and the accuracy dropped from $96\%$ to $50\%$ for the males and $92\%$ to $45\%$ for the females. The details of the leave-one-out experiment are discussed in the next subsection.

\subsection{\label{}Speaker Independent Cross-validation Experiment}
In this experimentation setting, the model was trained on $7$ actors and tested on the remaining actor with overall testing accuracy, recall, precision, and f1-score of $39\%$, $39\%$, $36\%$, and $35\%$, respectively, averaged over all actors as shown in Table \ref{our_results}. In the individual actor analysis, predictions on actor $3$ have, relatively, better results. As mentioned earlier, this below-par accuracy is due to a significantly different distribution of features for individual actors, as seen in Figure \ref{all_ser}a. The model fails on an unseen actor because of diversity in style to utter emotions. Training accuracy was observed to be $100\%$, even with extensive experimentation with reduction of layers, the addition of regularization, low learning rate, the testing accuracy did not improve, which shows that a simple deep neural network is not suitable for this variation of the experiment. We propose that complex models like LSTMs and transformers should be tested for better representation of heterogeneous distributions among speakers.

Moreover, our model can only identify anger and surprise emotions for all actors and perform well while predicting anger emotion as compared to others as shown in Figure \ref{all_ser}b. Boredom, happiness, and sadness emotions cannot be discriminated against for all actors. Disgust has the lowest overall accuracy for all speakers. We conclude that there exists heterogeneity among speakers and emotions collectively which is only natural considering diversity in emotion utterance for each individual, i.e., elicitation of similar emotions can vary speaker-wise.  

\begin{table*}
\caption{SER results for different variations against our proposed deep neural network model. }
\label{our_results}
\begin{tabular}{c @{\hskip 0.5in} c c c c}
\toprule
\textbf{Techniques/Evaluation Metrics} & \textbf{Accuracy} & \textbf{Precision} & \textbf{Recall} & \textbf{F1- score} \\ \midrule
Leave one out speaker                  & 0.39              & 0.39               & 0.36            & 0.35               \\ 
\textbf{Gender Analysis-Male}
&\textbf{0.96}                   &\textbf{0.96}                    &\textbf{0.96}                 &\textbf{0.96}                    \\ 
Gender analysis -Female                &0.92                   &0.92                    &0.92                 &0.92                    \\ 

\textbf{Random splits}                          &  \textbf{0.92}                 &\textbf{0.93}                    &\textbf{0.92}                 &\textbf{0.93}                    \\ \bottomrule
\Description{This table elaborates various experimentation setting used for testing SEMOUR in terms of accuracy, precision, recall and F1 score.}
\end{tabular}
\end{table*}

\section{Discussion}
In this section, we elaborate on the design and construction of our dataset and the results of the human annotation process. We also discuss the limitations of our current work in terms of dialect representation, machine-based model generalization, and speech spontaneity. 

\subsection{Dataset Construction Process}

This section summarizes the pursuit of gathering a high-quality database followed by the authors for the production or extension of any dataset. The first and foremost step is to select a language and study relevant published work to identify possible gaps in the resources available for that language. In light of our problem statement, a phonetically balanced script was designed enriched with words, phrases, and sentences to be elicited in diverse emotions. Rich vocabulary sentences in the neutral speech were available \cite{raza2009design} but not useful in our scenario as the vocabulary employed in these is not used in daily routine, hence a new script had to be designed to target the spoken Urdu language composed from frequently used words. The next step was to recruit actors by advertisement following a strict screening process based upon their language proficiency and performance experience. The script was distributed among the speakers for practice before recordings. A soundproof recording room was booked for actors to record their sessions. Actors with a performance lacking in the authenticity of emotions were asked to rerecord for better elicitation. Hence, recordings against each actor's emotion were procured and clipped according to the strict order and correct pronunciation of the script's instances.

Once the sound clips were ready, one-third of the repository was tagged by annotators to obtain human accuracy. A user-friendly application to aid annotation was designed to achieve this goal. Annotators were asked to tag discrete categorical and continuous attributes along with the authenticity of sound clips. Extensive experimentation to address accuracy and performance measures were performed along with comparative analysis for providing a fine benchmark for further explorations on the repository. Figure \ref{main_diagram} elaborates the aforementioned steps for procuring this dataset.

The authors would like to highlight that each step of acquisition posed diverse challenges. Instances of the script were modified continuously until a reasonable balance in phonemes and their frequencies were achieved as compared to the existing language corpus and most frequently used word list. Moreover, various re-takes were performed to ensure correct pronunciations and authentic emotion elicitation by actors. Post-processing of audio clips after successful recording sessions was indeed a tedious task. Gaps of silences were removed then clips were pruned and renamed according to instances of the script. During this process, a list of mistakes was maintained for the actors to elicit the mispronounced instances again. Experimentation on the dataset was only made possible once all the instances were correctly saved.

\subsection{Interpreting Human Subjects Study}

As discussed in section \ref{section_human_evaluation}
we used human subjects to evaluate one-third of the instances in SEMOUR. The goal of this study was to verify whether the expressions in these instances represent the corresponding emotion labels. Among the randomly chosen subset, $78\%$ of the instances were correctly classified by a human evaluator on average. We concluded that these $78\%$ instances contain enough information to detect a correlation between the acted expression and the corresponding emotion label. For the remaining $22\%$ instances, there are two possibilities. Either those instances were not uttered or recorded with the correct emotional expression, or, the instances contain the correct acted expression but it was missed by the human evaluator and they classified it incorrectly.
We designed an experiment to investigate the cause of these misclassifications by human evaluators as follows. We trained a simple Neural Network on the instances that were correctly classified by the evaluators. So, the model learned to distinguish the correct expression based on the labels of instances on which human evaluators agreed with dataset labels.
We tested this Neural Network model on the remaining instances that were either acted incorrectly by actors or misclassified which human evaluators.
The Neural Network achieved a test accuracy of $92\%$ on these instances. This indicated a significant correlation between the emotion-labels and the features of respective utterances. We conclude that these correlations may have been missed by the evaluators.

There is also a concern whether the uttered instances contain exaggerated expressions of emotions which may lead to a system that won't generalize to spontaneous speech. We asked our human subjects to tag whether a given acted expression was perceived as natural or exaggerated. Of the $9977$ sample instances for which we received a vote, $>84\%$ were reported to have a natural expression while the remaining $<16\%$ were tagged with a fake or exaggerated expression. We conclude that most of the acted expressions in SEMOUR successfully mimic the corresponding natural expressions in spontaneous speech.

\subsection{Limitations}

Semour is the first dataset of scripted emotional speech dataset for the Urdu Language recorded in eight complex emotions. Naturally, there are several limitations in this work that offer interesting avenues for improvement. Below, we discuss these limitations and provide future directions for this work.

Studies on speech recognition systems have noted several disadvantages to using acted speech as compared to data from spontaneous conversations in natural settings \cite{douglas2005multimodal,batliner2000desperately}. As the ultimate goal of a recognition system is to classify uncontrolled speech, a model trained on natural datasets is expected to generalize better. Similarly, the spontaneous speech also provides a context that is lacking in the acted counterpart \cite{cauldwell2000did}.
Despite these benefits, most of the works on speech emotion recognition, including SEMOUR, are based on datasets that use acted speech expressions as mentioned in \cite{swain2018databases}. The general drawback in building any potential datasets of spontaneous speech is that the audio samples are unlabelled by nature and depend on either the subjective opinion of the people tagging the dataset or some machine-assisted predictions to establish the ground truth. Secondly, the collection and cleaning process of spontaneous data requires considerably more effort \cite{douglas2003emotional}. For instance, the voice quality in the natural datasets has a high variance that results in a significant loss of accuracy, as studied in \cite{scherer2003vocal}. Therefore, one needs more advanced methods for noise removal. As one expects, natural datasets are also highly unbalanced for emotion classes. An overwhelming majority of the instances need to be discarded because they contain redundant information that is unproductive for model training.
The construction of a natural speech dataset while overcoming these challenges is an open problem. Meanwhile, there is ample evidence to suggest that acted speech expressions provide a good approximation to spontaneous speech \cite{jurgens2015effect}. Also, acted speech expressions are ideal to train on for certain applications. One such potential application is to understand the extent of any inherent emotional bias towards a community in movies. Since unobserved target utterances are acted expressions, it is a good idea to have a model train on acted instances.

Urdu has four dialects and several accents that depend on the demographics of the speaker \cite{dialects}. Due to a lack of resources and accessibility, utterances in SEMOUR are limited to a single \textit{(Urdu)} dialect spoken by the native people of Lahore. An extension covering rich dialect representation is required in this repository. Our current work also lacks the study of the impact of demographic effects on actors' emotion elicitation as well as taggers' human annotation.

Lastly, We report a modest 39\% accuracy on the leave-one-out experiment. We strongly believe this is due to the limitations of the simple NN that we use. Our main contribution is a dataset, and we have not invested efforts into a sophisticated model. The goal of the experiments is to provide a baseline for future works. Improving the accuracy of classification on unknown data using a deep model is an interesting future direction. Training a model to predict emotions for a previously unknown speaker is, indeed, a challenging problem. For example, a CNN-BLSTM based model on the IEMOCAP dataset reports 52\% accuracy for 6 emotions \cite{zhao2019speech}. This also highlights a problem regarding diversity in speaker profiles. Along with utilizing a deep model, learning better speaker embeddings might improve speaker-independent accuracy.

For future direction, we plan to diversify SEMOUR to cover all the major dialects as well as accents of the Urdu Language. With this goal in mind, we invite the research community to expand the dataset further to help create a universal speech emotion recognition system for Urdu. The dataset and other relevant metadata are available at \url{acoustics-lab.itu.edu.pk} through a request form. We allow the research community to download and independently extend and improve the dataset for non-commercial research purposes. We have also devised a system where the researchers can help improve the current version of the SEMOUR dataset by supplementing the recorded samples in different dialects and accents. The researchers can get the dataset and the script from \url{acoustics-lab.itu.edu.pk} and have their uttered collection appended in the current version after quality checks. We also intend to expand the current version on our own to develop a more generalized speech emotion recognition system.

\section{Conclusion}
In this work, we provide a first of its kind, gender, and phonetically balanced, scripted emotional speech dataset for the Urdu Language recorded by eight actors in eight complex emotions with $15,040$ unique instances. A uniformly selected one-third of the instances in SEMOUR are manually tagged and validated by resulting in human accuracy of $78\%$ and high inter-evaluator correlation scores. We also provide a $5$-layered neural network for speech emotion recognition task with variations of experiments on SEMOUR in comparison with classical machine learning techniques. Our model performs with an average accuracy of $90\%$ for a $10$-fold random split experiment. 

\begin{acks}

This work is partially supported by the Higher Education Commission (HEC), Pakistan under the National Center for Big Data and Cloud Computing funding for the Crime Investigation and Prevention Lab (CIPL) project at Information Technology University, Lahore.
We are thankful to the anonymous reviewers for their remarks and suggestions that helped significantly improve the quality of this paper.
We acknowledge the efforts of our volunteers including 
Sidra Shuja, Abbas Ahmad Khan, Abdullah Rao, Talha Riaz, Shawaiz Butt, Fatima Sultan, Naheed Bashir, Farrah Zaheer, Deborah Eric, Maryam Zaheer, Abdullah Zaheer, Anwar Said, Farooq Zaman, Fareed Ud Din Munawwar, Muhammad Junaid Ahmad, Taha Chohan, and Sufyan Khalid. 
We also thank the staff at ITU FM Radio 90.4 for their help in the recording process.
\end{acks}

\bibliographystyle{ACM-Reference-Format}
\bibliography{references.tex}

\end{document}